\documentclass[preprint,showpacs,superscriptaddress,nofootinbib]{revtex4}
\usepackage[dvips]{graphicx}
\usepackage{amsmath,amssymb}

\def\lromn#1{\uppercase\expandafter{\romannumeral#1}}

\def\av#1{\langle #1 \rangle}

\begin{document}

  \preprint{UT-HET 072}
  
  \title{
  Radiative type-I seesaw model with dark matter \\ 
  via U(1)$_{B-L}$ gauge symmetry breaking at future linear colliders} 
  \author{Shinya Kanemura}
  \email{kanemu@sci.u-toyama.ac.jp}
  \affiliation{
  Department of Physics,
  University of Toyama, Toyama 930-8555, Japan
  }
  \author{Takehiro Nabeshima}
  \email{nabe@jodo.sci.u-toyama.ac.jp},
  \affiliation{
  Department of Physics,
  University of Toyama, Toyama 930-8555, Japan
  }
  \author{Hiroaki Sugiyama} 
  \email{sugiyama@sci.u-toyama.ac.jp}
  \affiliation{
  Department of Physics,
  University of Toyama, Toyama 930-8555, Japan
  }
  \begin{abstract}
  We discuss phenomenology of the radiative seesaw model 
  in which spontaneous breaking of the U(1)$_{B-L}$ gauge symmetry at the TeV scale 
  gives the common origin for masses of neutrinos and dark matter~\cite{Kanemura:2011mw}.
  In this model, the stability of dark matter is realized by 
  the global U(1)$_{DM}$ symmetry which arises by the B$-$L charge assignment. 
  Right-handed neutrinos 
  obtain TeV scale Majorana masses at the tree level. 
  Dirac masses of neutrinos are generated via one-loop diagrams. 
  Consequently, tiny neutrino masses are generated at the two-loop level
  by the seesaw mechanism. 
  This model gives characteristic predictions, such as 
  light decayable right-handed neutrinos, 
  Dirac fermion dark matter and an extra heavy vector boson. 
  These new particles would be accessible at collider experiments
  because their masses are at the TeV scale. 
  The U(1)$_{B-L}$ vector boson may be found at the LHC, 
  while the other new particles could only be tested at future linear colliders. 
  We find that the dark matter can be observed at a linear collider 
  with $\sqrt{s}$=500~GeV and that light right-handed neutrinos can also be probed 
  with $\sqrt{s}$=1~TeV. 
  \end{abstract} 


\maketitle

\section{Introduction}

The standard model of elementary particle physics (SM) is a very successful model 
which explains ${\cal O}$(100)~GeV physics. 
However, we must consider beyond the SM because several phenomena such as 
neutrino oscillation~\cite{solar, atom, acc-app, acc-disapp, short-reac, long-reac}, 
dark matter (DM)~\cite{dark matter, WMAP} and baryon asymmetry of Universe~\cite{Sakharov:1967dj} 
cannot be explained by the SM. 

The neutrino oscillation can be explained by tiny neutrino masses. 
The simplest method to obtain these masses is 
the seesaw mechanism~\cite{seesaw} with right-handed Majorana neutrinos (the type-I seesaw scenario). 
In this mechanism, right-handed neutrino masses are ${\cal O}$($10^{15}$)~GeV 
for ${\cal O}$(1) neutrino Dirac Yukawa couplings. 
However, these heavy right-handed neutrinos would be difficult to detected at future experiments. 
On the other hand, if tiny neutrino masses are generated at the loop level, 
new particle masses can be ${\cal O}$(1)~TeV with sizable coupling constants. 
Moreover, this mechanism is often applied to the models with 
the $Z_2$ parity or an U(1) symmetry 
to forbid tree-level neutrino masses and 
to guarantee the stability of dark matter~\cite{
Gu:2007ug, Kanemura:2011jj, KNT, Ma, Ma:2008ba, AKS, Kanemura:2011mw, Kanemura:2012rj, Law:2012mj, 
Farzan:2012sa, Okada:2012np,Kanemura:2010bq,Suematsu:2010nd,Kanemura:2011vm,Lindner:2011it}. 
If dark matter is a Weakly Interacting Massive Particle (WIMP), 
its mass is naively estimated to be of 
${\cal O}$(100-1000)~GeV by the WMAP data~\cite{WMAP}. 
Such a mass scale for the dark matter may be naturally generated 
by the vacuum expectation value with respect to 
the spontaneous break down at the TeV-scale of 
an additional gauge symmetry. 

Let us consider 
the radiative seesaw model 
in which spontaneous breaking of the U(1)$_{B-L}$ gauge symmetry at the TeV scale 
gives the common origin for masses of neutrinos and dark matter~\cite{Kanemura:2011mw}.
In this model, the stability of dark matter is realized by 
the global U(1)$_{DM}$ symmetry which arises by B$-$L charge assignment. 
In this model, 
right-handed neutrinos 
obtain TeV-scale Majorana masses at the tree level, 
while Dirac masses of neutrinos are generated via one-loop diagrams 
where the dark matter is running in the loop. 
Consequently, tiny neutrino masses are generated at the two-loop level
by the seesaw mechanism. 
This model gives characteristic predictions, such as 
light decayable right-handed neutrinos, 
Dirac fermion WIMP dark matter and 
an extra heavy vector boson with the mass to be the TeV scale. 
These new particles would be accessible at collider experiments 
so that the model would be testable. 
For example, the LHC may be able to detect 
the U(1)$_{B-L}$ gauge boson ($Z'$) with the mass up to about 4~TeV 
with the U(1)$_{B-L}$ gauge coupling constant about 0.1~\cite{Z'B-L}. 
However, it would be challenging to test the other new particles in this model at the LHC 
unless the $Z'$ boson is sufficiently light, 
because they only couple to the $Z'$ boson and leptons 
so that their discovery strongly depends on the mass of the $Z'$ boson. 

In this paper, we discuss a possibility of testing 
this model at future electron-positron linear colliders~\cite{ILC, CLIC}. 
Differently from the hadron collider, 
new Dirac fermions can be produced directly from the $e^+e^-$ collision, 
whose sequential decays can contain production of right-handed neutrinos with the dark matter. 
The dark matter can also be directly produced 
in association with charged inert scalar bosons from leptons. 
We find that 
for the center-of-mass energy being $\sqrt{s} = 350$~GeV with the integrated luminosity of 1~ab$^{-1}$, 
the Dirac fermion dark matter can be detected as long as its mass is less than about 64~GeV. 
For the center-of-mass energy $\sqrt{s} = 500$~GeV with the same luminosity, 
the dark matter with the mass of ${\cal O}$(100)~GeV can be detected. 
Moreover, at the linear collider with $\sqrt{s} = 1$~TeV with 3 ab$^{-1}$ 
right-handed neutrinos can be tested. 
Therefore, this model can be identified by the combination of these future experiments. 

This paper is organized as follows. 
In the next section, we introduce the radiative type-I seesaw model. 
We discuss prospect at the LHC in Sec.3 and the ILC  in Sec.4.
Conclusions are given in Sec.5. 
\section{The radiative type-I seesaw model}
\begin{table}[t]
\begin{center}
\begin{tabular}
{|p{25mm}|@{\vrule width 1.8pt\ }p{15mm}|p{15mm}|p{15mm}|p{15mm}|p{15mm}|p{15mm}|}
   \hline
    Particles
     & $s^0$
     & $\eta$
     & $\Psi_{Ri}$
     & $\Psi_{Li}$
     & $\nu_{Ri}^{}$
     & $\sigma^0$
    \\ \noalign{\hrule height 1.8pt}
    SU(3)$_{\rm C}$
     & {\bf \underline{1}}
     & {\bf \underline{1}}
     & {\bf \underline{1}}
     & {\bf \underline{1}}
     & {\bf \underline{1}}
     & {\bf \underline{1}}
    \\ \hline
    SU(2)$_{\rm L}$
     & {\bf \underline{1}}
     & {\bf \underline{2}}
     & {\bf \underline{1}}
     & {\bf \underline{1}}
     & {\bf \underline{1}}
     & {\bf \underline{1}}
    \\ \hline
    U(1)$_{\rm Y}$
     & 0
     & $1/2$
     & 0
     & 0
     & 0
     & 0 
    \\ \hline
    U(1)$_{B-L}$
     & $1/2$
     & $1/2$
     & $-1/2$
     & $3/2$
     & $1$
     & $2$
    \\ \hline
\end{tabular}
\caption{New particles and their quantum charges.}
\label{table:particles}
\end{center}
\end{table}
We consider 
the TeV-scale seesaw model which explains tiny neutrino masses, the dark matter mass 
and the stability of the dark matter via the U(1)$_{B-L}$ gauge symmetry breaking~\cite{Kanemura:2011mw}. 
New particles and their gauge charges are shown in Table~\ref{table:particles}. 
Here, $s^0, \eta$ and $\sigma^0$ are complex scalar fields while 
$\nu_{Ri}^{}, \Psi_{Ri}$ and $\Psi_{Li}$ ($i$ = 1,2) are Wely fermions. 
This model has U(1)$_{B-L}$ anomalies but 
those anomalies can be cancelled by new fermions 
whose masses are larger than O(10)~TeV. 

Yukawa interactions are given by
\begin{eqnarray}
 {\cal L}_{\text{Yukawa}}
 &=&
 {\cal L}_{\text{SM-Yukawa}}
 -
 y_{Ri}^{}\,
 \overline{ (\nu_{Ri}^{})^c }\, \nu_{Ri}^{}\, (\sigma^0)^\ast
 -
 y_{\Psi_i}\,
 \overline{ \Psi_{Ri} }\, \Psi_{Li}\, (\sigma^0)^\ast
 \nonumber\\
 &&\hspace*{-2mm}
{}-
 (y_{3})_{ij}\,
 \overline{ (\nu_{Ri}^{})^c }\, \Psi_{Rj}\, (s^0)^\ast
 -
 h_{ij}\,
 \overline{ \Psi_{Li} }\, \nu_{Rj}^{}\, s^0
 \nonumber\\
 &&{}\hspace*{-2mm}
 -
  f_{\ell i}\,
 \overline{ L_{L\ell} }\, \Psi_{Ri}\, i\sigma_2\, \eta^\ast
 + \text{h.c.} ,
 \nonumber \\
 \label{eq:Lagrangian}
\end{eqnarray}
where $ {\cal L}_{\text{SM-Yukawa}}$ denotes Yukawa interactions of the SM and 
$L_{L_\ell}$ is the left-handed lepton doublet of $\ell\ (\ell= e,\mu,\tau)$ flavor. 
Masses of $\nu_R^{}, \Psi_R$ and $\Psi_L$ are forbidden by the U(1)$_{B-L}$ gauge symmetry. 
We take a basis
where Yukawa matrices $y_R$ and $y_\Psi$ are diagonalized
such that their real positive eigenvalues satisfy
$y_{R1}^{} \leq y_{R2}^{}$ and $y_{\Psi_1}^{} \leq y_{\Psi_2}^{}$.

Scalar potential is given by 
\begin{eqnarray}
 V(\Phi,s.\eta,\sigma)
 &=&
 -\mu_{\phi}^2\Phi^{\dagger} \Phi
 +
 \mu_s^2 |s^0|^2
 +
 \mu_{\eta}^2\eta^{\dagger} \eta
 -
 \mu_{\sigma}^2 |\sigma^0|^2
 \nonumber\\
 &&{}
 +
 \lambda_\phi \left(\Phi^{\dagger} \Phi\right)^2
 +
 \lambda_s |s^0|^4
 +
 \lambda_\eta \left(\eta^{\dagger} \eta\right)^2
 +
 \lambda_\sigma |\sigma^0|^4
 \nonumber\\
 &&{}
 +
 \lambda_{s\eta} |s^0|^2 \eta^{\dagger} \eta
 +
 \lambda_{s\phi} |s^0|^2 \Phi^{\dagger} \Phi
 +
 \lambda_{\phi\phi} (\eta^{\dagger}\eta) (\Phi^{\dagger} \Phi)
 +
 \lambda_{\eta\phi} (\eta^{\dagger} \Phi) (\Phi^{\dagger}\eta)
 \nonumber\\
 &&{}
 +
 \lambda_{s\sigma} |s^0|^2 |\sigma^0|^2
 +
 \lambda_{\sigma\eta} |\sigma^0|^2 \eta^{\dagger} \eta
 +
 \lambda_{\sigma\phi} |\sigma^0|^2 \Phi^{\dagger} \Phi
 +\left( \mu_3^{}\, s^0\, \eta^{\dagger}\, \Phi + \text{h.c.}\right) ,
 \label{eq:potencial}
 \nonumber \\
 \end{eqnarray}
where $\Phi$ is the SM Higgs doublet field 
and $\mu_\phi^2$, $\mu_\sigma^2$, $\mu_s^2$, and $\mu_\eta^2$ 
are positive values. 
In our model, there appears global U(1) symmetry (we name it the U(1)$_{DM}$) 
under which 
$s^0, \eta, \Psi_{Ri}$ and $\Psi_{Li}$ have the same charge. 
The U(1)$_{DM}$ symmetry stabilizes the dark matter. 
After the breakings of the U(1)$_{B-L}$ and electroweak symmetries, 
$\sigma^0$ and $\phi^0$ obtain vevs 
$v_\sigma$~[$= \sqrt{2} \av{\sigma^0}$] and 
$v$~[$=\sqrt{2} \av{\phi^0} \simeq 246\,$GeV], respectively.  
The U(1)$_{B-L}$ gauge boson $Z^\prime$ acquires its mass as $m_{Z^\prime} = 2 g_{B-L}^{} v_\sigma$,
where $g_{B-L}^{}$ denotes the gauge coupling constant of the U(1)$_{B-L}$.

Right-handed neutrinos $\nu_{Ri}^{}$ obtain Majorana masses
$M_{Ri}$~[$= \sqrt{2} y_{Ri} v_\sigma$] 
while $\Psi_{Ri}$ and $\Psi_{Li}$ for each $i$
become a Dirac fermion $\Psi_i$
with its mass $M_{\Psi_i}$~[$= y_{\Psi_i} v_\sigma/\sqrt{2}$].
Since the global $U(1)_{DM}$ is not broken by $v_\sigma$,
the lightest $U(1)_{DM}$-charged particle is stable.
We assume $\Psi_1$ to be lightest $U(1)_{DM}$ particle 
and it becomes a candidate for the dark matter.

After symmetry breaking with $v_\sigma$ and $v$,
mass eigenstates of two CP-even scalars
and their mixing angle $\alpha$
are given by
\begin{eqnarray}
\begin{pmatrix}
 h^0\\
 H^0
\end{pmatrix}
=
 \begin{pmatrix}
  \cos\alpha & -\sin\alpha\\
  \sin\alpha & \cos\alpha
 \end{pmatrix}
 \begin{pmatrix}
  \phi^0_r\\
  \sigma^0_r
 \end{pmatrix} , \quad
%
\sin{2\alpha}
=
 \frac{ 2\lambda_{\sigma\phi} v v_\sigma }{ m_{H^0}^2 - m_{h^0}^2 } ,
\end{eqnarray}
where $\sigma^0 = (v_\sigma + \sigma^0_r + iz_\sigma)/\sqrt{2}$
and $\phi^0 = (v + \phi^0_r + iz_\phi)/\sqrt{2}$,
and 
$z_\phi$ and $z_\sigma$ are Nambu-Goldstone bosons 
absorbed by $Z$ and $Z^\prime$, respectively. 
Masses of $h^0$ and $H^0$ are defined by 
\begin{eqnarray}
m_{h^0}^2
&=&
 \lambda_\phi v^2 + \lambda_\sigma v_\sigma^2
 -\sqrt{
   \left(
    \lambda_\phi v^2 - \lambda_\sigma v_\sigma^2
   \right)^2
   + \lambda_{\sigma\phi}^2 v^2 v_\sigma^2 }\, ,
 \nonumber\\
%
m_{H^0}^2
&=&
 \lambda_\phi v^2 + \lambda_\sigma v_\sigma^2
 +\sqrt{
   \left(
    \lambda_\phi v^2 - \lambda_\sigma v_\sigma^2
   \right)^2
   + \lambda_{\sigma\phi}^2 v^2 v_\sigma^2 }\, .
\end{eqnarray}
On the other hand,
since $s^0$ and $\eta^0$ are $U(1)_{DM}$-charged particles,
they are not mixed with $\sigma^0$ and $\phi^0$. 
Mass eigenstates of these $U(1)_{DM}$-charged scalars
and their mixing angle $\theta$
are obtained as
\begin{eqnarray}
\begin{pmatrix}
 s^0_1\\
 s^0_2
\end{pmatrix}
=
 \begin{pmatrix}
  \cos\theta & -\sin\theta\\
  \sin\theta & \cos\theta
 \end{pmatrix}
 \begin{pmatrix}
  \eta^0\\
  s^0
 \end{pmatrix} , \quad
%
\sin{2\theta}
=
 \frac{ \sqrt{2} \mu_3^{} v }{ m_{s^0_2}^2 - m_{s^0_1}^2 } .
\end{eqnarray}
Mass eigenvalues $m_{s^0_1}$ and $m_{s^0_2}$
of these neutral complex scalars
are defined by
\begin{eqnarray}
m_{s^0_1}^2
&=&
 \frac{1}{2}
 \left(
  m_\eta^2 + m_s^2
  -\sqrt{ \left( m_\eta^2 - m_s^2 \right)^2 + 2 \mu_3^2 v^2 }
 \right),
\nonumber\\
%
m_{s^0_2}^2
&=&
 \frac{1}{2}
 \left(
  m_\eta^2 + m_s^2
  +\sqrt{ \left( m_\eta^2 - m_s^2 \right)^2 + 2 \mu_3^2 v^2 }
 \right) ,
\end{eqnarray}
where
$m_s^2
= \mu_s^2 + \lambda_{s\phi} v_{\phi}^2/2 + \lambda_{s\sigma} v_{\sigma}^2/2$
and
$m_{\eta}^2
= \mu_{\eta}^2
 + \left( \lambda_{\phi\phi} + \lambda_{\eta\phi} \right) v_{\phi}^2/2
 + \lambda_{\sigma\eta} v_{\sigma}^2/2$.
Finally,
the mass of the charged scalar $\eta^\pm$ is given by 
\begin{eqnarray}
m_{\eta^{\pm}}^2
&=&
 \mu_\eta^2 +\lambda_{\phi\phi} \frac{v^2}{\,2\,}
 + \lambda_{\sigma\eta} \frac{v_{\sigma}^2}{2} .
 \label{eq:m_etapm}
\end{eqnarray}

In Table~\ref{table:parameter}, 
we show an examples for the parameter set 
which satisfies current experimental bounds such as 
neutrino oscillation, $\mu\to e\gamma$ constraint, 
relic abundance of dark matter and dark matter direct detection~\cite{Kanemura:2011mw}. 
We define $M_{R}$ as a common mass of two $\nu_{Ri}^{}$ ($i=1, 2$). 
\begin{table}[t]
\begin{center}
\begin{tabular}{|c|@{\vrule width 1.8pt}c|c|}
\hline
 $f_{\ell i}$
  &\rule[0mm]{0mm}{15mm}
   $
    \begin{pmatrix}
     0.0757 \ & 0.0445\\
     0.01 \ & -0.0123\\
     -0.141 \ & -0.0101
    \end{pmatrix}
   $
   \\[10mm]
\hline
 $h_{ij}$
  &\rule[0mm]{0mm}{10mm}
   $
    \begin{pmatrix}
     -0.131 \ & 0.1\\
     0.1 \ & 0.1\\
    \end{pmatrix}
   $
   \\[5mm]
\hline
 $(y_3)_{ij}$
  &\rule[0mm]{0mm}{10mm}
   $
    \begin{pmatrix}
     0.0152 \ & 0.0152\\
     0.0152 \ & 0.0152\\
    \end{pmatrix}
   $
   \\[5mm]
\hline
 \ $M_R \equiv M_{R1} = M_{R2}$ \
  & $250~{\rm GeV}$\\
\hline
 $\{ M_{\Psi_1},\ M_{\Psi_2} \}$
  & $\{ 57.5~{\rm GeV},\ 800~{\rm GeV} \}$\\
\hline
 $\{ m_{h^0}^{},\ m_{H^0}^{},\ \cos\alpha \}$
  & \ $\{ 125~{\rm GeV},\ 130~{\rm GeV},\ 1/\sqrt{2} \}$ \\
\hline
 $\{ m_{s^0_1},\ m_{s^0_2},\ \cos\theta \}$
  & $\{ 200~{\rm GeV},\ 300~{\rm GeV},\ 0.05 \}$\\
\hline
 $m_{\eta^\pm}$
  & $280~{\rm GeV}$\\
\hline
 $g_{B-L}^{}$
  & $0.1$\\
\hline
 $m_{Z'}^{}$
  & $4000~{\rm GeV}$\\
\hline
\end{tabular}
\caption{The parameter set which satisfies current experimental bounds.
Our analysis is insensitive to the Higgs boson masses $m_{h^0}^{}$ and $m_{H^0}^{}$.
}
\label{table:parameter}
\end{center}
\end{table}
\section{Physics at the LHC} 
We first consider testability of the model at the LHC. 
We discuss physics of Higgs bosons ($h^0$ and $H^0$),
the U(1)$_{B-L}$ gauge boson Z', 
and the right-handed neutrinos $\nu_R^{}$. 
\subsection{Detectability of h and H}
Our model predicts two SM-like CP-even Higgs bosons $h^0$ and $H^0$ 
because a large mixing angle between $h^0$ and $H^0$ is 
required to be consistent with the dark matter abundance~\cite{Kanemura:2011mw}. 

In the SM, the Higgs boson mass 
except for the small region between 122.5~GeV and 127~GeV 
has already been excluded by the recent LHC results~\cite{atlas, cms}. 
However, in our model, 
we can read off from the SM Higgs boson search results that 
wider regions for $m_{h^0}^{}$ and $m_{H^0}^{}$ remain allowed; i.e., 
between 110~GeV and 130~GeV, 
because production cross sections for $h^0$ and $H^0$ are reduced by the factor 
with maximal mixing ($1/2$) from the SM predictions. 

When more data will be accumulated at the LHC, 
the allowed region will be smaller. 
Then, only the case with approximately degenerated masses may be allowed, 
where the two resonances are overlapped and look like one resonance of only one SM Higgs boson 
depending on the resolution of detectors. 
If the mass difference is larger than about 1~GeV, 
the two resonances can be separated at the LHC~\cite{atlas, cms}. 
Even if the mass difference is smaller, 
they could be separated at the ILC 
where the mass can be reconstructed with much better accuracy with the error of 
$\Delta m \sim 50$~MeV~\cite{Abe:2010aa}.

\subsection{Detectability of Z'}
Detectability of the U(1)$_{B-L}$ gauge boson Z' at the LHC 
has been studied in several papers~\cite{Z'B-L, Z'B-L2, Erler:2011ud}. 
The most efficient process of production of the Z' at the LHC is the Drell-Yan mode 
$q\overline{q}\to Z'$ where $q$ is a quark in the proton~\cite{Z'B-L, Z'B-L2, Erler:2011ud}. 
The Z' boson dose not mix with the SM Z boson at the tree level. 
Therefore, branching ratios of the Z' boson decay are 
proportional to U(1)$_{B-L}$ charges of daughter particles. 
The $q\overline{q}\to Z'\to \mu^-\mu^+$ process is useful to detect the Z' at the LHC. 
By using this process, the Z' boson can be detected at the 3 $\sigma$ Confidence Level (C.L.) 
for $m_{Z'}^{} \lesssim 4$~TeV with U(1)$_{B-L}$ gauge coupling constant $g_{B-L}^{}\simeq 0.1$
at the LHC with $\sqrt{s}=14$~TeV and 100~fb$^{-1}$~\cite{Z'B-L}. 
Essentially, the Z' boson can be distinguished from the other gauge bosons by 
checking the ratio of $Z'\to \mu^-\mu^+$ to $Z'\to b\overline{b}$.

\subsection{Detectability of right-handed neutrino}
The right-handed neutrinos would be produced via the decay of the Z' at the LHC: 
\begin{eqnarray} 
q\overline{q}\to Z'\to\nu_{R}^{}\overline{\nu_{R}^{}} .
 \label{eq:nR_detect_LHC}
\end{eqnarray}
In our model, 
if the decay of right-handed neutrinos into the dark matter and a SM singlet scalar $s_1^{0}$ 
is kinematically forbidden, 
the possible decay of the right-handed neutrinos is $\nu_{R}^{}\to \ell^{\pm}W^{\mp}$. 
In this case, detectability of the right-handed neutrinos at the LHC 
has been studied in the simplest U(1)$_{B-L}$ model in Refs.~\cite{Z'B-L, Z'B-L2}. 
For example, the right-handed neutrinos can be detected at 26.3 $\sigma$ C.L. 
with $M_{R}$ = 200~GeV, $m_{Z'}^{}$ = 1.5~TeV and $g_{B-L}$ = 0.2 
at the LHC with $\sqrt{s}=14$~TeV and 100~fb$^{-1}$~\cite{Z'B-L}. 
\section{Physics at the ILC}
We here consider testability of our radiative type-I seesaw model at the ILC. 
It is difficult to detect signals of the model at the LHC 
if the U(1)$_{B-L}$ gauge boson Z' is too heavy to be produced. 
Even for the case, the ILC can test the dark matter, right-handed neutrinos 
and the scalar sector. 
\subsection{Detectability of dark matter} 
\begin{figure}[t]
 \begin{tabular}{cc}
  \begin{minipage}{0.5\hsize}
   \begin{center}
    \scalebox{0.15}{\includegraphics{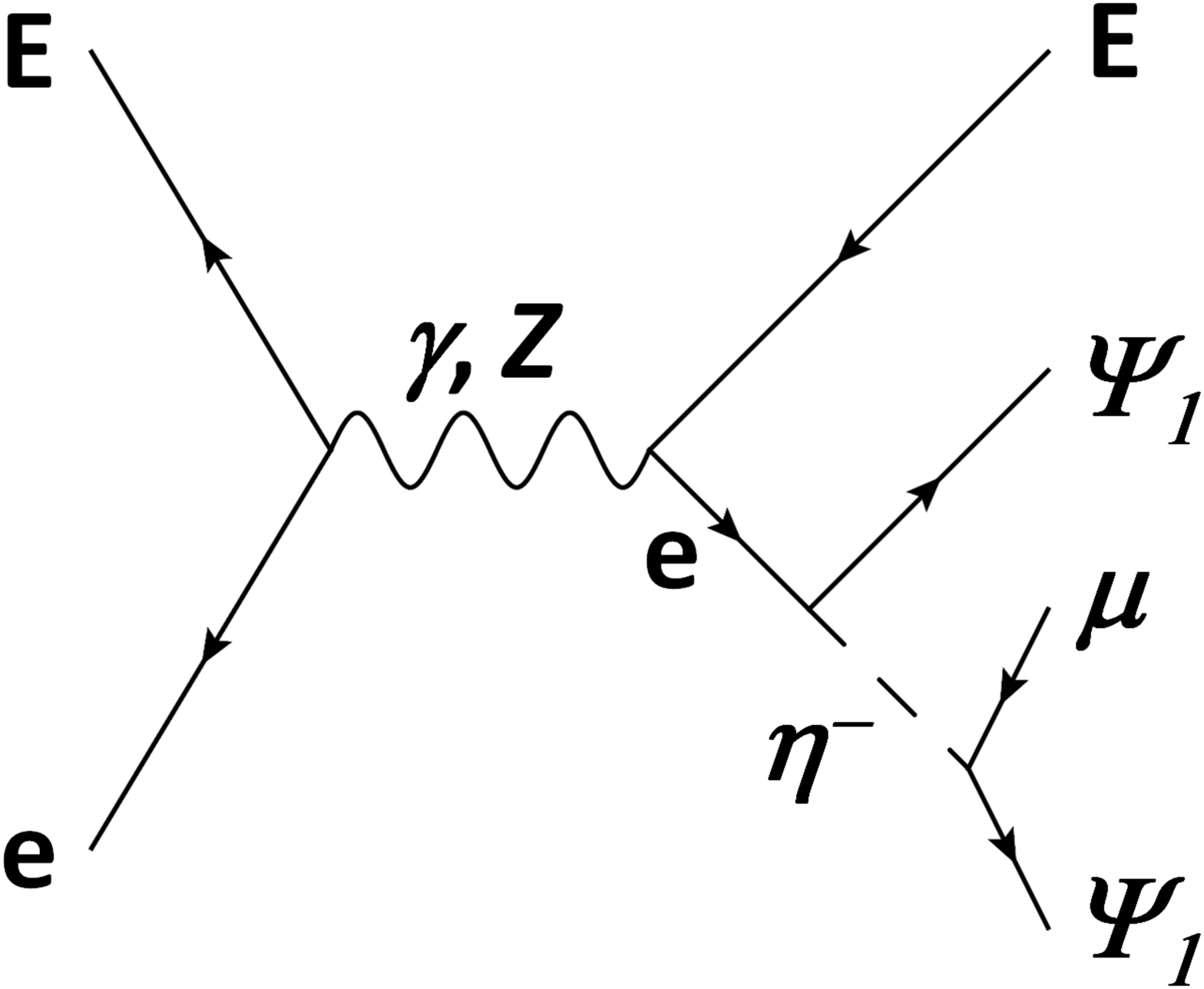}}
   \end{center}
  \end{minipage}
  \begin{minipage}{0.5\hsize} 
   \begin{center}
    \scalebox{0.15}{\includegraphics{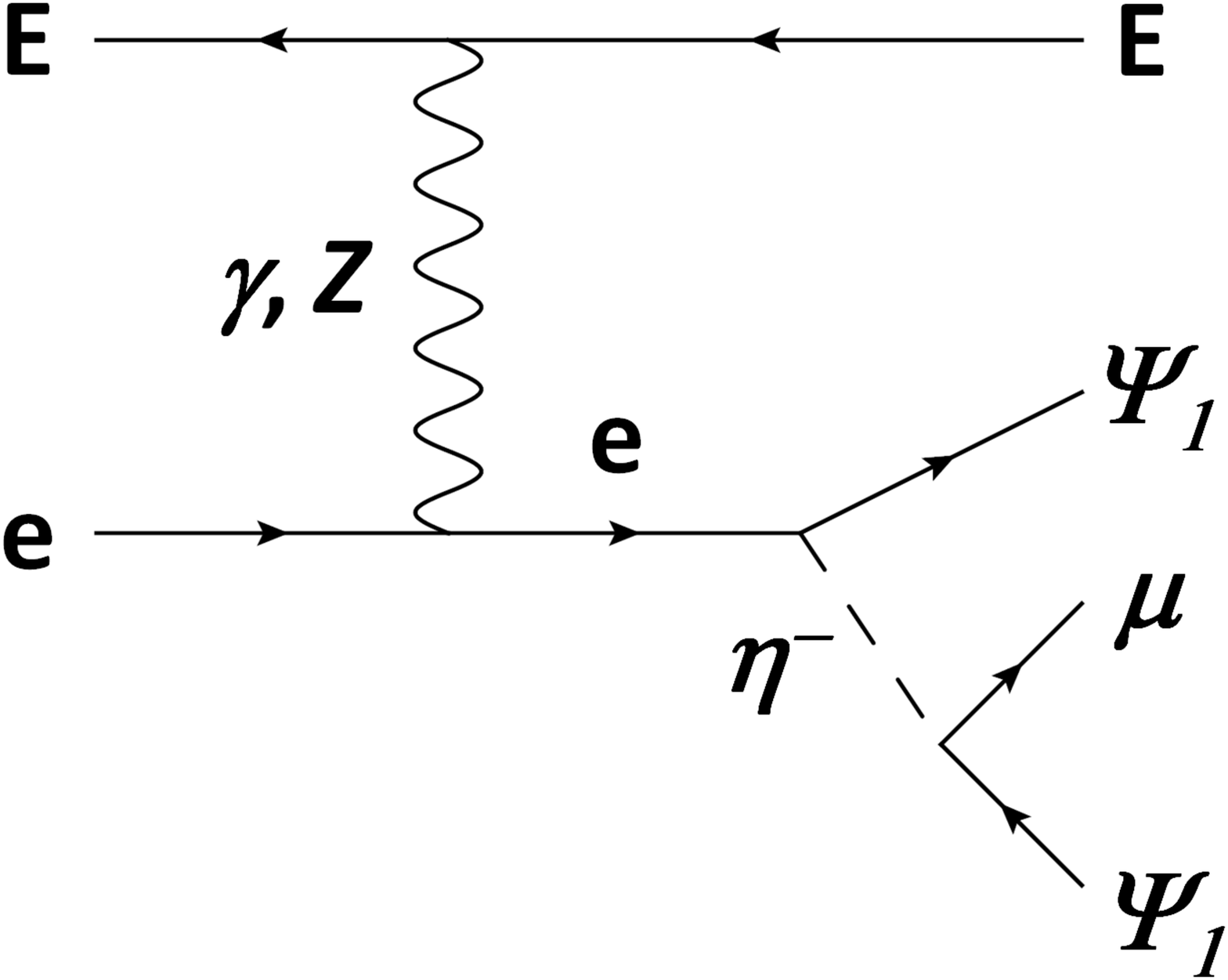}}
   \end{center}
  \end{minipage}
 \end{tabular}
  \caption{Feynman diagrams for the signal process of the dark matter.}
  \label{fig:DMsig}
\end{figure}

In our model, 
the mass of the dark matter is about a half of the SM-like Higgs boson mass ($m_{h^0}^{}/2$ or $m_{H^0}^{}/2$)
to satisfy the WMAP data. 
The dark matter directly couples to charged leptons. 
We concentrate on the process 
$e^+e^-\to e^+\Psi_1\eta^-\to e^+\mu^-\Psi_1\overline{\Psi_1}$ 
, where $\eta^-$ is an on-shell state, depicted in Fig.~\ref{fig:DMsig}. 
The dark matter can be detected as missing particles 
by using the energy momentum conservation for the process.

We assume $\sqrt{s} = 350$~GeV and $\sqrt{s} = 500$~GeV at the ILC. 
The outgoing electron tends to be emitted to forward and backward directions 
because this is a $t$-channel process. 
Therefore, the detectability of the leptons near the beam line is important. 
We assume that detectable regions of the scatter angle ($\theta_\ell$) of the emitted lepton are 
given for $\sqrt{s} = 350$~GeV and $\sqrt{s} = 500$~GeV by 
$|\cos (\theta_\ell)|<0.99955$ and $|\cos (\theta_\ell)|<0.9998$, respectively~\cite{Abramowicz:2010bg}. 

The background processes are 
charged lepton flavor changing ones with missing momenta such as 
$e^+e^-\to e^+W^-\nu_e^{}\to e^+\mu^-\nu_e^{}\overline{\nu_{\mu}^{}}$. 
In order to increase signal significance, 
we impose kinematical cuts for the case with $\sqrt{s} = 350$~GeV as follows: 
\begin{eqnarray}
 0.96
 &<&
 |\cos (\theta_\ell)|
 <
 0.99955,
\ \ \ 
 \cos(\theta_{e\mu})
 <
 0.
 \ \ \ 
 M_{e\mu}
 <
 80~{\rm GeV},
\ \ \ 
\nonumber \\
 E_{e}
 &<&
 30~{\rm GeV},
 \ \ \ 
 E_{\mu}
 >
 100~{\rm GeV},
\ \ \ 
 120~{\rm GeV}
 <
 M_{miss}
 <
 190~{\rm GeV},
  \label{eq:350cut}
\end{eqnarray} 
where $\theta_{e\mu}, M_{e\mu}, E_{\ell}$ and $M_{miss}$ are 
the angle between electron and muon, invariant mass of electron and muon, 
energy of $\ell$ and missing invariant mass, respectively. 
The following cuts are imposed for the case with $\sqrt{s}=500$~GeV: 
\begin{eqnarray}
 0.98&<&
 |\cos (\theta_e)|
 <
 0.9998,
\ \ \ 
 |\cos (\theta_\mu)|
 <
 0.8,
\ \ \  
 |\cos(\theta_{e\mu})|
 <
 0.8
\nonumber \\
 E_{e}
 &<&
 120~{\rm GeV},
 \ \ \ 
  E_{\mu}
 >
 80~{\rm GeV},
\ \ \ 
 M_{miss}
 >
 120~{\rm GeV}.
 \label{eq:500cut}
\end{eqnarray}

\begin{figure}[t]
 \begin{tabular}{cc}
  \begin{minipage}{0.5\hsize}
   \begin{center}
    \scalebox{0.4}{\includegraphics[angle=-90]{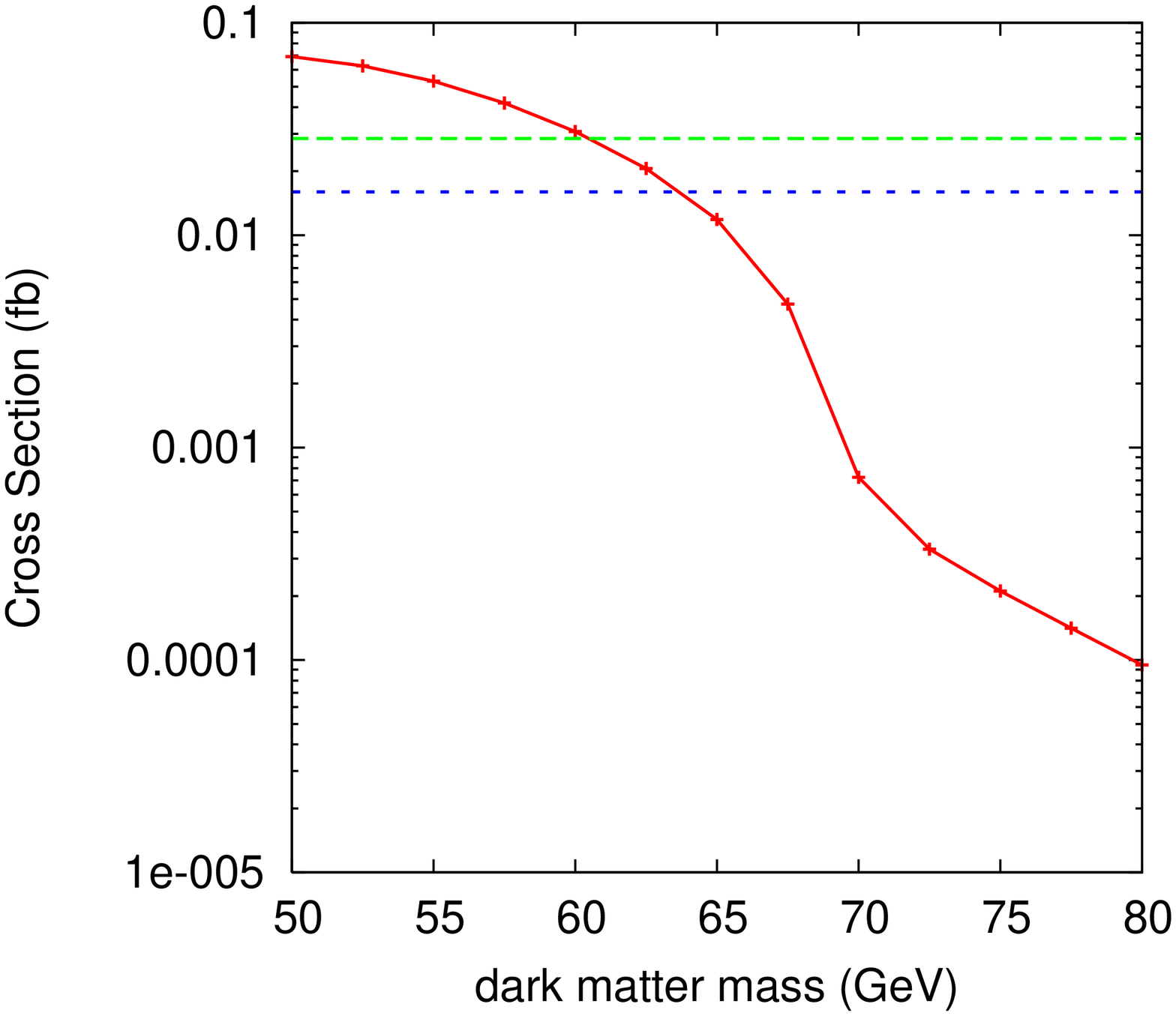}}
   \end{center}
  \end{minipage}
  \begin{minipage}{0.5\hsize} 
   \begin{center}
    \scalebox{0.4}{\includegraphics[angle=-90]{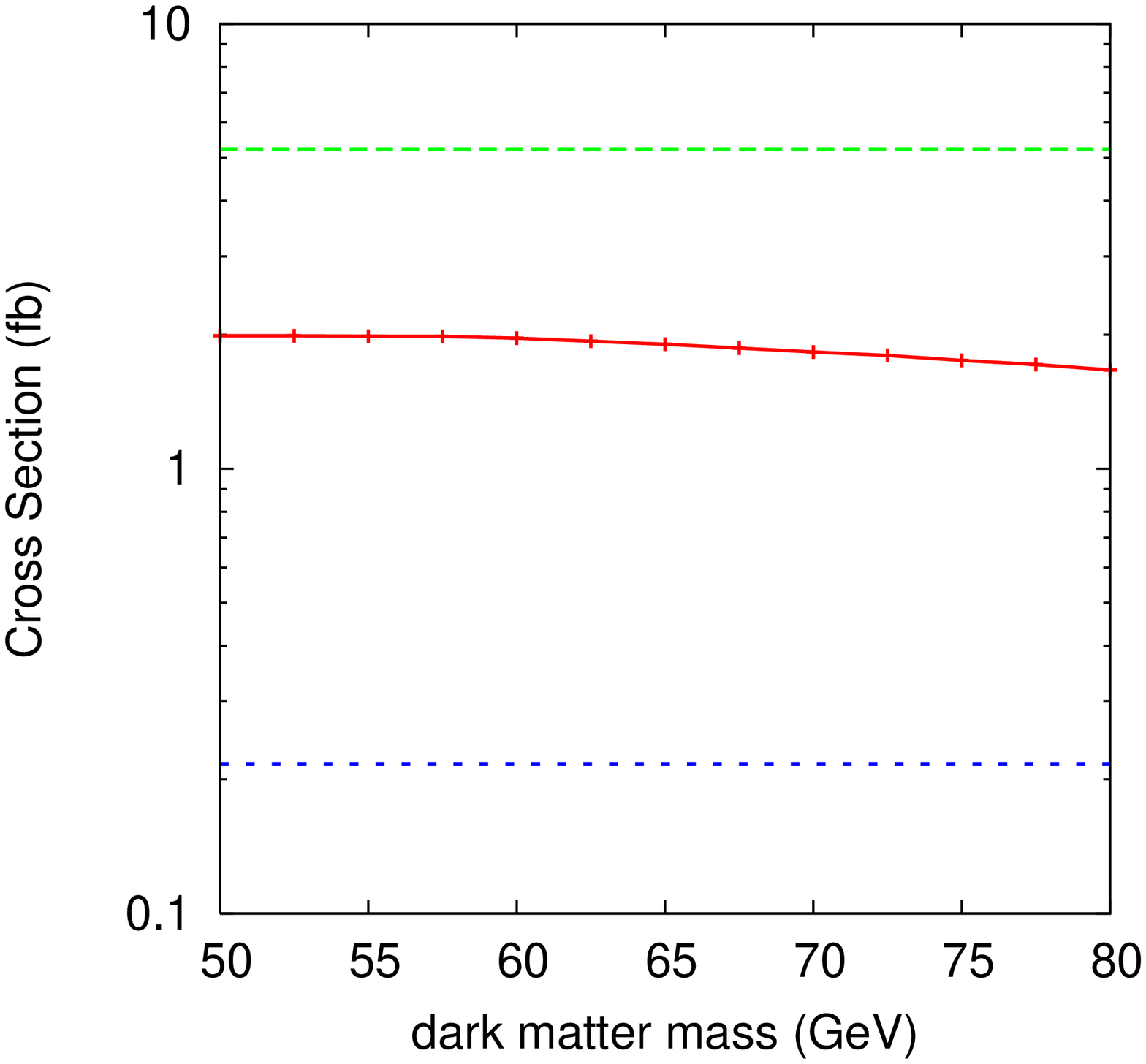}}
   \end{center}
  \end{minipage}
 \end{tabular}
 \caption{The cross section of signal of the dark matter 
  at the ILC with $\sqrt{s}=350$~GeV (left) and $\sqrt{s}=500$~GeV (right).
  The red (solid) line and the green (dashed) lines are signal and background cross section, respectively. 
  The blue (dotted) line means the limit that the signal significance $N_S/\sqrt{N_B} = 3$ 
  with 1~ab$^{-1}$. 
}
 \label{fig:DMXSAC}
\end{figure}

We show the detectability of the dark matter after applying all cuts for 
$\sqrt{s} = 350$~GeV and $\sqrt{s} = 500$~GeV 
in left and right panels of Fig.~\ref{fig:DMXSAC}, respectively. 
The vertical axis is the cross section 
and the horizontal axis is the dark matter mass. 
All parameters but $M_{\Psi_1}^{}$ are set to the values in Table~\ref{table:parameter}.
The red (solid) line and the green (dashed) line represent signal and background cross sections, respectively. 
The blue (dotted) line means the cross section for which the signal significance 
($N_S/\sqrt{N_B}$ where $N_S$ and $N_B$ are number of events for signal and backgrounds) 
is 3 with 1~ab$^{-1}$. 
The left panel of Fig.~\ref{fig:DMXSAC} show that the dark matter can be detected 
at the ILC with $\sqrt{s} = 350$~GeV and 1~ab$^{-1}$ for $M_{\Psi_1} \lesssim$ 64~GeV at 3$\sigma$ C.L. 
In the right panel of Fig.~\ref{fig:DMXSAC}, we find that the dark matter can be detected 
for $M_{\Psi_1} \lesssim$ 80~GeV 
at the ILC with $\sqrt{s} = 500$~GeV and 1~ab$^{-1}$ at more than 3$\sigma$ C.L. 

\begin{figure}[t]
 \begin{tabular}{cc}
  \begin{minipage}{0.5\hsize}
   \begin{center}
    \scalebox{0.4}{\includegraphics[angle=-90]{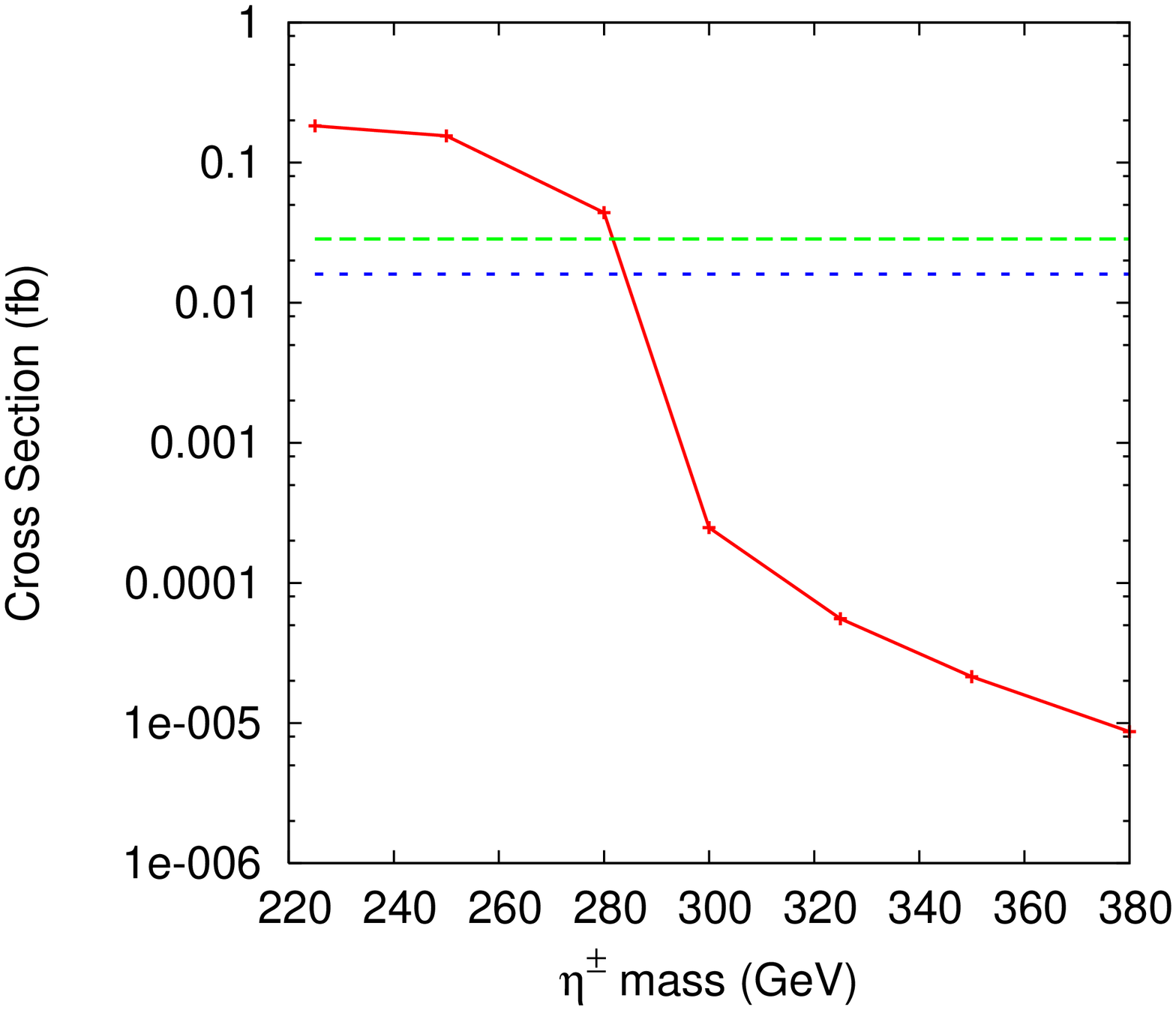}}
   \end{center}
  \end{minipage}
  \begin{minipage}{0.5\hsize} 
   \begin{center}
    \scalebox{0.4}{\includegraphics[angle=-90]{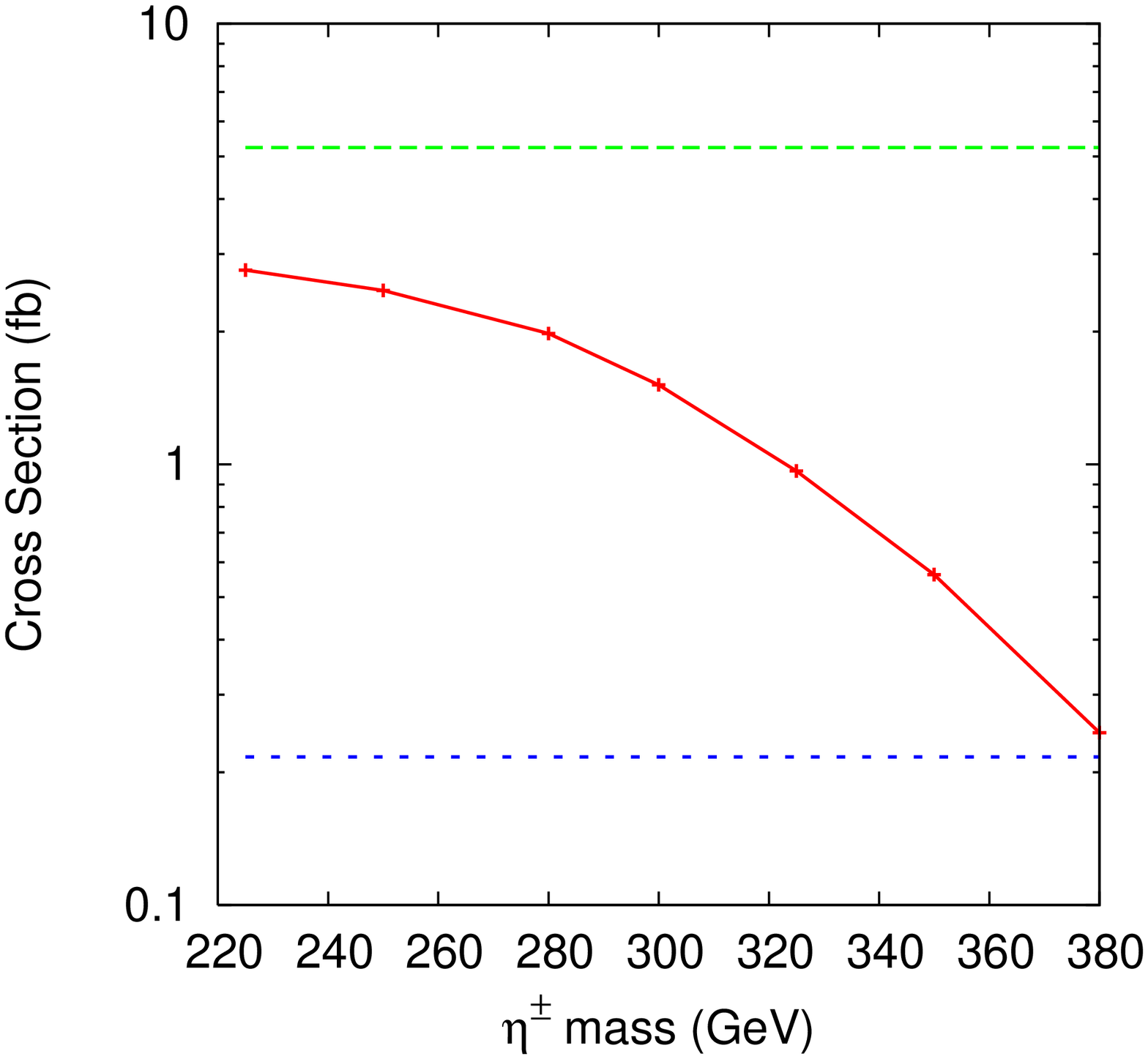}}
   \end{center}
  \end{minipage}
 \end{tabular}
 \caption{The dependence of the signal cross section on mass of the $\eta^{\pm}$ 
  at the ILC with $\sqrt{s}=350$~GeV (left) and $\sqrt{s}=500$~GeV (right).
  Meanings of all of lines are the same as those in Fig.~\ref{fig:DMXSAC}. }
 \label{fig:DMXSACeta}
\end{figure}

The cross section of the signal process depends on the mass of $\eta^{\pm}$. 
In left and right panels of Fig.~\ref{fig:DMXSACeta}, 
we show the dependence of the signal cross section on $m_{\eta^{\pm}}^{}$ 
after applying all cuts for 
$\sqrt{s} = 350$~GeV and $\sqrt{s} = 500$~GeV, respectively. 
Parameters but $m_{\eta^{\pm}}^{}$ are set on the values in Table~\ref{table:parameter}.
Meanings of all of lines the same as those in Fig.~\ref{fig:DMXSAC}. 
The dark matter would be detected at 
the ILC with $\sqrt{s} = 350$~GeV and $\sqrt{s} = 500$~GeV for 
$m_{\eta^{\pm}}^{} \lesssim 280$~GeV and $m_{\eta^{\pm}}^{} \lesssim 380$~GeV, 
respectively.

If the decay of the SM-like Higgs bosons into the dark matter is kinematically allowed, 
its branching ratios are ${\cal O}$(1)\%. 
In this case, dark matter would be detected at the ILC with $\sqrt{s} = 350$~GeV~\cite{Schumacher:2003ss}. 
Furthermore, direct detection experiments of dark matter are very important for our model. 
Since the annihilation cross section of $\Psi_1$ is proportional to $v_{\sigma}^{-2}$, 
the dark matter abundance becomes inconsistent with the WMAP data~\cite{WMAP} if $v_\sigma^{} \gtrsim$ 16~TeV. 
Therefore, the spin independent scattering cross section between dark matter and neutron via $Z'$ boson 
is more than 2$\times 10^{-47}$~cm$^{-2}$ ,
and this signal could be detected at the XENON1T~\cite{Aprile:2012zx}. 
If dark matter mass measured at direct detection experiments is consistent with $m_{\Psi_1}^{}$ measured at the ILC,
the $\Psi_1$ would be identified as dark matter.
In this case, the lepton flavor changing process of 
$e^+e^-\to e^+\Psi_1\eta^-\to e^+\mu^-\Psi_1\overline{\Psi_1}$ 
indicates that 
the dark matter is directly coupled to charged leptons. 
Moreover, if we utilize the beam polarization, 
it could be checked that the dark matter only couples with left-handed leptons. 
That kind of the dark matter would contribute to neutrino mass generation mechanism. 

Notice that $e^+e^-\to e^+\Psi_1\eta^-\to e^+\tau^-\Psi_1\overline{\Psi_1}$ process 
would be also useful 
to detect the dark matter because the $f_{\mu 1}$ coupling constant 
is constrained by $\mu\to e\gamma$ 
while the $f_{\tau 1}$ coupling constant is not strictly constrained. 
Therefore, the cross section for the $\tau$ mode tends to be larger than the one for the $\mu$ mode. 
If the dark matter can be detected in both $\mu$ and $\tau$ modes, 
we test the ratio between $f_{\mu 1}$ and $f_{\tau 1}$. 

\subsection{Detectability of right-handed neutrinos}
\begin{figure}[t]
 \begin{center}
  \scalebox{0.3}{\includegraphics{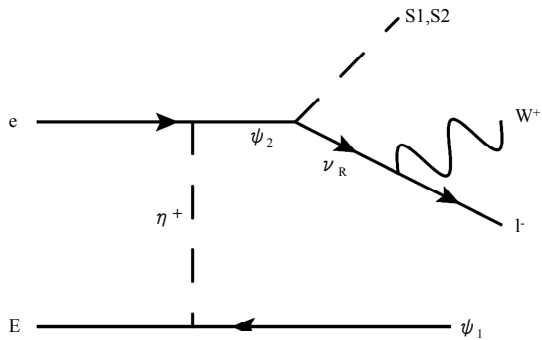}}
  \caption{Feynman diagram of the signal process of right-handed neutrinos.}
  \label{fig:RNsig}
 \end{center}
\end{figure}

Next, we consider detectability of right-handed neutrinos at the ILC. 
In our model, neutrino Dirac Yukawa and SM gauge coupling constants of 
right-handed neutrinos are not large. 
However, right-handed neutrinos couple to U(1)$_{DM}$ charged particles such as $\Psi_{2}$ 
with sizable coupling constants of ${\cal O}$(0.1). 
In this case, right-handed neutrinos can be produced via $\Psi_{2}$.
We concentrate on the process, 
$e^+e^-\to \Psi_1\overline{\Psi_2}\to \Psi_1\overline{\nu_R^{}}s_1^0\to\Psi_1W^-\ell^+s_1^0$ 
where $\ell$ is electron and muon, depicted in Fig.~\ref{fig:RNsig}. 
We use the two-jet mode of the W boson decay while $s_1^0$ decays invisibly into 
$\Psi_1$ and left-handed neutrinos. 

In our choice of the parameters of the model, 
the center of mass energy around 1~TeV is necessary to produce $\Psi_2$ at the ILC. 
We here consider the case with $\sqrt{s} = 1$~TeV 
for right-handed neutrino detection via $\Psi_2\to s_1^0\nu_R^{}$. 
Main background process is $e^+e^-\to W^-\ell^+\nu_i\gamma$ 
with photon missing events. 
We impose the following kinematical cuts 
in order to increase signal significance: 
\begin{eqnarray}
 &&
 |\cos (\theta_\ell)|
 <
 0.95,
 \ \ \ 
 200~{\rm GeV}
 <
 M_{miss}
 <
 600~{\rm GeV},
 \nonumber \\
 &&0.9999416
 <
 |\cos (\theta_\gamma)|, 
 \nonumber \\
 &&240~{\rm GeV}
 <
 M_{{\rm W}\ell}
 <
 260~{\rm GeV},
 \ \ \ 
 E_{\ell} <
 300~{\rm GeV}, 
 \nonumber \\
 &&300~{\rm GeV}
 <
 E_{{\rm W}\ell}
 <
 600~{\rm GeV}, 
 \label{eq:allNR}
\end{eqnarray}
 where $\theta_x$ represent the scattering angle of the particle $x$, 
 and $E_{{\rm w}\ell}$ and $M_{{\rm w}\ell}$ are 
 the energy and the invariant mass between the W boson and $\ell$, respectively. 
 
\begin{figure}[t]
 \begin{center}
  \scalebox{0.5}{\includegraphics[angle=-90]{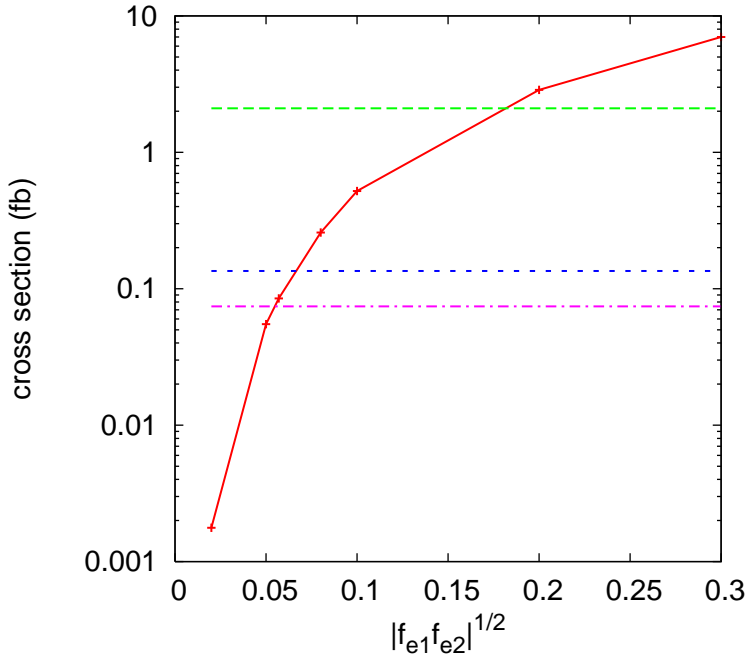}}
  \caption{Right-handed neutrino detectability at the $\sqrt{s}=1$~TeV ILC. 
  The red (solid) line and the green (dashed) line are signal and background cross section 
  times BR($W\to jj$). 
  The blue (dotted) and magenta (dash-dotted) lines mean the limit that the signal significance ($N_S/\sqrt{N_B}$) 
  are three with 1~ab$^{-1}$ and 3~ab$^{-1}$, respectively. 
} 
  \label{fig:RMsig}
 \end{center}
\end{figure}

In Fig.~\ref{fig:RMsig}, 
we show the detectability of right-handed neutrinos after applying all cuts. 
The horizontal axis is $\sqrt{f_{e1}f_{e2}}$ 
and the other parameters are set on values in Table~\ref{table:parameter}. 
The red (solid) line and the green (dashed) line are signal and background cross section times BR($W\to jj$), 
respectively. 
Cross sections for which the signal significance ($N_S/\sqrt{N_B}$) becomes 3 are presented with
the blue (dotted) line for 1~ab$^{-1}$ and the magenta (dash-dotted) line for 3~ab$^{-1}$. 
In Fig.~\ref{fig:RMsig}, we find that right-handed neutrinos can be detected at the 3$\sigma$ C.L. 
at the ILC with $\sqrt{s} = 1$~TeV and 3~ab$^{-1}$ in our parameter set. 
If $M_{R} > M_{\Psi_1} + m_{s_1^0}^{}$, 
right-handed neutrinos can decay into U(1)$_{DM}$ charged particles. 
In this case, $\nu_{R}^{} \to \Psi_1s_1^0$ is the main decay mode of right-handed neutrinos. 
Therefore most of right-handed neutrinos decay into U(1)$_{DM}$ charged particles invisibly, 
and it is difficult to detect right-handed neutrinos. 

\subsection{Scalar sector}
Two SM-like CP-even Higgs bosons ($h^0$ and $H^0$) with the large mixing 
are predicted in our model. 
If the 125~GeV boson discovered at the LHC 
is identified to the SM-like Higgs boson, 
it means that both of masses of $h^0$ and $H^0$ are close to this signal mass
in our model. 
The energy resolution for the Higgs boson is expected to be 
better than about 50~MeV at the ILC~\cite{Abe:2010aa}. 
Therefore, two CP-even Higgs bosons would be separated at the ILC. 
Notice that there is no CP-odd scalar boson without U(1)$_{DM}$ charge in this model.
On the other hand, charged scalar bosons $\eta^{\pm}$ dominantly decay into 
left-handed charged lepton and dark matter. 
This decay process is similar to slepton decay in super symmetric models. 
If $\eta^{\pm}$ are produced in pair at TeV-scale colliders, 
$m_{\eta_{\pm}}^{}$ would be reconstructed using the maximum $M_{T2}$ value~\cite{Lester:1999tx}
or lepton energy distribution~\cite{Kawabata:2011gz}.

\section{Conclusions}

We have investigated testability of the TeV-scale radiative type-I seesaw model 
at TeV-scale colliders, 
which explains tiny neutrino masses, 
the fermionic dark matter mass and the stability of the dark matter by 
introducing the U(1)$_{B-L}$ gauge symmetry. 
We have found that 
the dark matter could be detected with the mass up to about 64~GeV 
with the integrated luminosity 1 ab$^{-1}$ at the ILC with $\sqrt{s} = 350$~GeV. 
The allowed region of the dark matter mass 
would be probed at the ILC with $\sqrt{s} = 500$~GeV. 
Since the value of $v_\sigma^{}$ is less than about 16~TeV for consistency with the WMAP data, 
dark matter direct detection experiments also could cover this region. 
Right-handed neutrinos could be discovered 
at the ILC with $\sqrt{s} = 1$~TeV and 1 ab$^{-1}$ 
if the invisible decay $\nu_R^{}\to \Psi_1 s_1^0$ is kinematically forbidden. 
Therefore, the model has sufficient possibilities to be distinguished from the other models 
by combination with dark matter search experiments and 
TeV-scale colliders such as the LHC and the ILC. 

\begin{acknowledgments}
This work of S.K. was supported in part by Grant-in-Aid for Scientific Research, Nos. 22244031
, 23104006 and 24340046. 
The work of H.S. was supported in part by the Grant-in-Aid for Young Scientists (B)
No. 23740210.
The work of T.N. was
supported in part by 
the Japan Society for the Promotion of Science as a research fellow (DC2).
\end{acknowledgments}

\end{document}